\documentclass[aps,prd,twocolumn,eqsecnum,showpacs,amsmath]{revtex4}
\usepackage[dvips]{color,graphicx}
\usepackage{amsfonts}
\usepackage{amssymb}
\usepackage{latexsym}

\newcommand{\dalm}{\kern1pt\vbox{\hrule height 0.9pt\hbox{\vrule width
0.9pt\hskip 2.5pt\vbox{\vskip 5.5pt}\hskip 3pt\vrule width 0.3pt}\hrule height
0.3pt}\kern1pt}
\newcommand{\ma}[1]{\mbox{$\mathcal{#1}$}}

\newcommand{\lw}[1]{\smash{\lower2.ex\hbox{#1}}}
\newtheorem{The}{Theorem}

\begin{document}


\title{Matter without matter: novel Kaluza-Klein spacetime in Einstein-Gauss-Bonnet gravity} 
\author{
$^{1,2,3}$Hideki Maeda\footnote{Electronic address:hideki@gravity.phys.waseda.ac.jp}
and
$^{4}$Naresh Dadhich\footnote{Electronic address:nkd@iucaa.ernet.in}}

\affiliation{
$^{1}$Graduate School of Science and Engineering, Waseda University, Tokyo 169-8555, Japan\\
$^{2}$Department of Physics, Rikkyo University, Tokyo 171-8501, Japan\\
$^{3}$Department of Physics, International Christian University, 3-10-2 Osawa, Mitaka-shi, Tokyo 181-8585, Japan\\
$^{4}$Inter-University Centre for Astronomy \& Astrophysics, Post Bag 4, Pune~411~007, India\\
}
\date{\today}

\begin{abstract}
We consider Einstein-Gauss-Bonnet gravity in $n(\ge 6)$-dimensional Kaluza-Klein spacetime ${\ma M}^{4} \times {\ma K}^{n-4}$, where ${\ma K}^{n-4}$ is the Einstein space with negative curvature.
In the case where ${\ma K}^{n-4}$ is the space of negative constant curvature, we have recently obtained a new static black-hole solution (Phys. Rev. D {\bf 74}, 021501(R) (2006), hep-th/0605031) which is a pure gravitational creation including Maxwell field in four-dimensional vacuum spacetime. 
The solution has been generalized to make it radially radiate null radiation representing gravitational creation of charged null dust.
The same class of solutions though exists in spacetime ${\ma M}^{d} \times {\ma K}^{n-d}$ for $d=3,4$, however the gravitational creation of the Maxwell field is achieved only for $d=4$. 
Also, Gauss-Bonnet effect could be brought down to ${\ma M}^d$ only for $d=4$.
Further some new exact solutions are obtained including its analogue in Taub-NUT spacetime on ${\ma M}^4$ for $d=4$. 
\end{abstract}


\maketitle

\section{Introduction}

Kaluza-Klein theory was a novel way of unifying Maxwell field and gravity~\cite{KK,ow1997}. 
It was shown that five-dimensional vacuum spacetime when appropriately projected onto four dimensions could incorporate Maxwell field. 
It was indeed the pioneering attempt of consideration of dimension higher than usual four. 
It was soon realized that the theory was not renormalizable, yet however it had remained smoldering all through in the background. 
With the advent of string theory, higher dimensions have now become the order of the day. 
In this paper, we wish to follow Kaluza-Klein spirit in studying Einstein-Gauss-Bonnet gravity.

In the previous paper~\cite{md2006}, we considered an $n(\ge 6)$-dimensional spacetime with Kaluza-Klein split with the topology of ${\ma M}^4 \times {\ma K}^{n-4}$, where ${\ma K}^{n-4}$ is the $(n-4)$-dimensional space of constant curvature with the constant warp factor $r_0$. 
It turns out that the vacuum Einstein-Gauss-Bonnet equation with a cosmological constant $\Lambda$ gets correspondingly split up into two parts~\cite{md2006}. The four-dimensional part is like the vacuum Einstein equation with a cosmological constant redefined while the extra dimensional part gives a scalar constraint equation. 
The constraint does not permit presence of a mass point and hence the vacuum solution reduces to a dS/AdS spacetime in four dimensions. 
However there exists an exceptional case in which four-dimensional equation could be made completely vacuous by prescribing $r_0$ and $\Lambda$ in terms of Gauss-Bonnet parameter $\alpha$ and the constant curvature of extra dimensions. 
Then $4$-metric is entirely determined by the scalar constraint given by the extra dimensional equation and it admits the general solution in the Schwarzschild gauge~\cite{md2006}. 

The solution represents a black hole with two horizons and ${\ma M}^4$ asymptotically approximates to Reissner-Nortstr\"om-anti-de~Sitter (RN-AdS) spacetime for positive $\alpha$ in spite of absence of Maxwell field. 
Note that the scalar constraint had prohibited presence of any matter, yet however there had come about a black hole with parameters which asymptotically resemble mass and Maxwell charge. This is what we call the pure gravitational creation, ``matter without matter'', produced by curvature of extra dimensional space through its linkage with $\alpha$ and $\Lambda$. It is purely a classical demonstration of reciprocity 
between matter and gravity (curvature) which is the first such example after the classic case of Kaluza-Klein. However in string theory, there occurs a similar situation in AdS/CFT correspondence~\cite{mal} in which at the boundary of AdS spacetime (gravity) resides conformal field theory of matter.
It would be of interest to examine this setting for the general case, ${\ma M}^{d} \times {\ma K}^{n-d}$ as well as for generalization to creation of other matter from curvature. 
This is what we precisely intend to do in the following.

The strong motivation for higher dimensions greater than usual four however comes from superstring/M-theory~\cite{superstring,Lukas}. 
However, one of us has recently argued that higher dimension is required even for complete description of classical dynamics of gravity~\cite{dad}. 
The action for gravity in higher dimensions is given by  Lovelock polynomial~\cite{lovelock} which is the most general action constructed from Riemann curvature giving rise to quasi-linear (highest order of differentiation occurring linearly) equation of motion. 
For $n\le 4$, it reduces to Einstein-Hilbert action with $\Lambda$. 
The next quadratic term in the action is called Gauss-Bonnet term which should be included for description of gravity for spacetime with dimension greater than four. 
It also arises in the low-energy limit of heterotic superstring theory as the higher curvature correction to general relativity~\cite{Gross}. 
Although Gauss-Bonnet term does not contribute to the field equation in four-dimensional spacetime, Kaluza-Klein decomposition of the higher-dimensional spacetime brings its contribution down to ${\ma M}^4$.

The Gauss-Bonnet term has nontrivial consequences in four-dimensional AdS gravity at the level of the conserved currents for the theory, in which the Gauss-Bonnet term regularizes the conserved quantities defined through the Noether theorem, canceling the usual divergences at radial infinity~\cite{olea2005}.
The Gauss-Bonnet coupling constant $\alpha$ is then fixed by demanding that the spacetime has constant curvature at the boundary.
Also, it was shown that the Gauss-Bonnet term regularizes the Euclidean action for four-dimensional AdS gravity and gives the correct entropy for a number of solutions~\cite{acotz2000}.

In this paper, we show that gravitational creation not only of Maxwell field but also of null dust.
The paper is organized as follows. 
In Section~II, we set up the general framework for Einstein-Gauss-Bonnet equation for the product spacetime and show that the same class of solutions (as for $d=4$) exists in spacetime $\ma M^{d} \times \ma K^{n-d}$ only for $d \le 4$.
In Section~III, we show that $d=4$ case is special in the sense that gravitational creation of Maxwell field is achieved as well as Gauss-Bonnet effects appear explicitly on ${\ma M}^d$.
We also generalize our solution to its analogues in Vaidya-like and Taub-NUT-like spacetimes on ${\ma M}^4$.
In Section~IV, we consider the general Lovelock case and argue that the solutions are generic to Lovelock gravity.
Section~V is devoted to discussion and conclusions.
A class of the Bohm metric with Einstein-space condition and its generalization with negative curvature are presented in Appendix A.   
The calculational setup for tensor decomposition is given in Appendix B.
Throughout this paper we use units such that $c=1$. 
As for notation we follow~\cite{Gravitation}. 
The Greek indices run $\mu=0,1, \cdots, n-1$. 

\section{Kaluza-Klein decomposition of basic equations}
We write action for $n$-dimensional spacetime, 
\begin{equation} 
\label{action}
S=\int d^nx\sqrt{-g}\biggl[\frac{1}{2\kappa_n^2}(R-2\Lambda+\alpha{L}_{GB}) \biggr]+S_{\rm matter},
\end{equation}
where $R$ and $\Lambda$ are $n$-dimensional Ricci scalar and the cosmological constant, respectively. 
Further $\kappa_n\equiv\sqrt{8\pi G_n}$, where $G_n$ is $n$-dimensional gravitational constant and $\alpha$ is the Gauss-Bonnet coupling constant.
The Gauss-Bonnet term ${L}_{GB}$ is combination of squares of Ricci scalar, Ricci tensor $R_{\mu\nu}$, and Riemann tensor $R^\mu_{~~\nu\rho\sigma}$ as
\begin{equation}
{L}_{GB} \equiv R^2-4R_{\mu\nu}R^{\mu\nu}+R_{\mu\nu\rho\sigma}R^{\mu\nu\rho\sigma}.
\end{equation}

This type of action is derived in the low-energy limit of heterotic superstring theory~\cite{Gross}.
In that case, $\alpha$ is identified with the inverse string tension and is positive definite. 
We shall therefore take $\alpha \ge 0$.

The gravitational equation following from the action (\ref{action}) is given by 
\begin{equation}
{\ma G}^\mu_{~~\nu} \equiv {G}^\mu_{~~\nu} +\alpha {H}^\mu_{~~\nu} +\Lambda \delta^\mu_{~~\nu}=\kappa_n^2 {T}^\mu_{~~\nu}, \label{beq}
\end{equation}
where 
\begin{eqnarray}
{G}_{\mu\nu}&\equiv&R_{\mu\nu}-{1\over 2}g_{\mu\nu}R,\\
{H}_{\mu\nu}&\equiv&2\Bigl[RR_{\mu\nu}-2R_{\mu\alpha}R^\alpha_{~\nu}-2R^{\alpha\beta}R_{\mu\alpha\nu\beta} \nonumber \\
&&~~~~+R_{\mu}^{~\alpha\beta\gamma}R_{\nu\alpha\beta\gamma}\Bigr]
-{1\over 2}g_{\mu\nu}{L}_{GB}.\label{def-H}
\end{eqnarray}
$T_{\mu\nu}$ is the energy-momentum tensor of the matter field derived from $S_{\rm matter}$ in the action (\ref{action}). 
It is noted that Gauss-Bonnet term makes no contribution in the field equations, i.e. $H_{\mu\nu} \equiv 0$, for $n \le 4$.

We consider $n$-dimensional Kaluza-Klein vacuum spacetime ($T_{\mu\nu}\equiv 0$) which is locally homeomorphic to ${\ma M}^{d} \times {\ma K}^{n-d}$ with the metric $g_{\mu\nu}=\mbox{diag}(g_{AB},r_0^2\gamma_{ab})$, $A,B = 0, \cdots, d-1;~a,b = d, \cdots, n-1$. 
Here $g_{AB}$ is an arbitrary Lorentz metric on ${\ma M}^d$, $r_0$ is a constant and $\gamma_{ab}$ is the unit metric on the $(n-d)$-dimensional Einstein space ${\ma K}^{n-d}$.

The $(n-d)$-dimensional Einstein space satisfies
\begin{eqnarray}
\overset{(n-d)}{R}{}_{abde}=\overset{(n-d)}{C}{}_{abde}+{\bar k}(\gamma_{ad}\gamma_{be}-\gamma_{ae}\gamma_{bd}),
\end{eqnarray}
where $\overset{(n-d)}{C}_{abde}$ is the Weyl tensor and ${\bar k}=\pm 1,0$.
The superscript $(n-d)$ means the geometrical quantity on ${\cal K}^{n-d}$.
If the Weyl tensor is identically zero, ${\ma K}^{n-d}$ is a space of constant curvature.
The Riemann tensor is contracted to give
\begin{eqnarray}
\overset{(n-d)}{R}{}_{ab}&=&{\bar k}(n-d-1)\gamma_{ab},\\
\overset{(n-d)}{R}&=&{\bar k}(n-d)(n-d-1).
\end{eqnarray}  

Now we assume the {\it Einstein-space condition} originally introduced by Dotti and Gleiser for $d=2$~\cite{dg2005,dg}:
\begin{eqnarray}
\overset{(n-d)}{C}{}^{abde}\overset{(n-d)}{C}{}_{fbde}&=&\Theta \delta^a_{~~f}. \label{hc}
\end{eqnarray}  
$\Theta$ must be constant by the consistency with the identity $H^{a\nu}_{~~;\nu}=0$.
The Weyl tensor vanishes identically in three or less dimensions, so that $\Theta \equiv 0$ holds for $n \le d+3$.
In~\cite{dg2005}, Dotti and Gleiser gave a unique example with ${\bar k}=1$ satisfying this condition in the class of the Bohm metric~\cite{bohm1998,ghp2003}.
Its metric and $\Theta$ are given by
\begin{eqnarray}
\gamma_{ab}dx^adx^b&=&d\rho^2+a(\rho)^2d\Omega_{m-1}^2+b(\rho)^2d\Omega_m^2, \label{bohm-s} \\
a(\rho)&=&\sqrt{\frac{m-1}{2m-1}}\sin\biggl(\sqrt{\frac{2m-1}{m-1}}\rho\biggl), \\
b(\rho)&=&\sqrt{\frac{2m-1}{m-1}}, \\
\Theta&=&\frac{2m(2m-1)}{m-1},
\end{eqnarray}  
where $d\Omega_p^2$ is the line element of a unit $p$-sphere and $m \ge 3$ holds~\cite{dg2005}. 
This metric has the positive curvature.
Here we have $2m=n-d$, so that $m \ge 3$ implies $n \ge d+6$.
This metric can be generalized to that with negative curvature (${\bar k}=-1$):
\begin{eqnarray}
\gamma_{ab}dx^adx^b&=&d\rho^2+a(\rho)^2d\Xi_{m-1}^2+b(\rho)^2d\Xi_m^2, \label{bohm-h} \\
a(\rho)&=&\sqrt{\frac{m-1}{2m-1}}\cosh\biggl(\sqrt{\frac{2m-1}{m-1}}\rho\biggl), \\
b(\rho)&=&\sqrt{\frac{2m-1}{m-1}}, \\
\Theta&=&\frac{2m(2m-1)}{m-1},
\end{eqnarray}  
where $d\Xi_p^2$ is the line element of a unit $p$-hyperboloid and $m \ge 3$ holds. 
The derivation of these metrics is presented in appendix~A.

Then ${\ma G}^\mu_{~~\nu}$ for the metric $g_{\mu\nu}=\mbox{diag}(g_{AB},r_0^2\gamma_{ab})$ gets decomposed as follows:
\begin{widetext}
\begin{eqnarray}
{\ma G}^A_{~~B}&=&\biggl[1+\frac{2{\bar k}\alpha(n-d)(n-d-1)}{r_0^2}\biggl]\overset{(d)}{G}{}^A_{~B}+\alpha\overset{(d)}{H}{}^A_{~B} \nonumber \\
&&+\biggl[\Lambda-\frac{{\bar k}(n-d)(n-d-1)}{2r_0^2}-\frac{{\bar k}^2\alpha(n-d)(n-d-1)(n-d-2)(n-d-3)}{2r_0^4}-\frac{(n-d)\alpha}{2r_0^4}\Theta\biggl] \delta^A_{~B},\label{dec1} \\
{\ma G}^a_{~~b}&=&\delta^a_{~~b}\biggl[-\frac12\overset{(d)}{R}+\Lambda-\frac{(n-d-1)(n-d-2){\bar k}}{2r_0^2} \nonumber \\
&&-\alpha\biggl\{\frac{{\bar k}(n-d-1)(n-d-2)}{r_0^2}\overset{(d)}{R}+\frac12 \overset{(d)}{L}_{GB} \nonumber \\
&&+\frac{(n-d-1)(n-d-2)(n-d-3)(n-d-4){\bar k}^2}{2r_0^4}+\frac{n-d-4}{2r_0^4}\Theta\biggl\}\biggl],\label{dec2}
\end{eqnarray}
\end{widetext}
where the superscript $(d)$ means the geometrical quantity on ${\ma M}^d$.  
(See Appendix B for derivation.)
The vacuum equation ${\ma G}^A_{~~B}=0$ is a tensorial equation on ${\ma M}^d$, while ${\ma G}^a_{~~b}=0$ is a scalar equation on ${\ma M}^{n-d}$ and works as a constraint.

 For $d=2$, we have $\overset{(2)}{G}{}^A_{~B}=\overset{(2)}{H}{}^A_{~B}=\overset{(2)}{L}_{GB} \equiv 0$.
Then, vacuum equations ${\ma G}^A_{~~B}=0$ and ${\ma G}^a_{~~b}=0$ give
\begin{eqnarray}
r_0^2&=&\frac{(n-2)(n-3)}{4\Lambda}\biggl({\bar k}\pm\sqrt{{\bar k}^2+B}\biggl), \label{Nariai1}\\
B&\equiv&\frac{8\alpha\Lambda[{\bar k}^2(n-3)(n-4)(n-5)+\Theta]}{(n-2)(n-3)^2},\\
\overset{(2)}{R}&=&\frac{2r_0^2}{r_0^2+2\alpha{\bar k}(n-3)(n-4)} \nonumber \\
&&~~~~~\times\biggl(\frac{4\Lambda}{n-2}-\frac{{\bar k}(n-3)}{r_0^2}\biggl) \label{Nariai2}
\end{eqnarray}  
for $n \ge 4$.
The parameters $\Lambda$, $\alpha$, ${\bar k}$, $n$, and $\Theta$ are taken for $r_0^2$ to be real and positive.
Eq.~(\ref{Nariai2}) implies that ${\ma M}^2$ is a two-dimensional constant curvature spacetime, which could be flat or (A)dS spacetime corresponding to $\overset{(2)}{R}=0$ or $\overset{(2)}{R}(<)>0$. 
In general relativity ($\alpha=0$), ${\bar k}\Lambda>0$ must be satisfied for $r_0^2$ to be real and positive.
The $n$-dimensional Nariai solution corresponds to the case with ${\bar k}=+1$ and $\Lambda>0$ with topology $dS^2 \times {\ma S}^{n-2}$, while the $n$-dimensional anti-Nariai solution corresponds to ${\bar k}=-1$ and $\Lambda<0$ with topology $AdS^2 \times {\ma H}^{n-2}$.
In Einstein-Gauss-Bonnet gravity, on the other hand, there is a variety of possible topology depending on the parameters.
The Nariai solution in Einstein-Gauss-Bonnet gravity with $\Theta=0$ was investigated in~\cite{Lorenz-Petzold1987a}.
Hereafter, we only consider the case with $d \ge 3$.

From the decomposition (\ref{dec1}) and (\ref{dec2}), the following well-known results in general relativity ($\alpha=0$) readily follow.
\begin{The} 
\label{the:1}
Let us consider $n(\ge d+1)$-dimensional vacuum Einstein equation without $\Lambda$.
Then, the metric $g_{\mu\nu}=\mbox{diag}(g_{AB},r_0^2\gamma_{ab})$ with ${\bar k}=0$ and arbitrary $r_0^2$ is the solution if $g_{AB}$ is the solution of the $d$-dimensional vacuum Einstein equation without $\Lambda$.
\end{The}
\begin{The} 
\label{the:2}
Let us consider $n(\ge d+2)$-dimensional vacuum Einstein equation with a non-zero cosmological constant $\Lambda$.
Then, the metric $g_{\mu\nu}=\mbox{diag}(g_{AB},r_0^2\gamma_{ab})$ with ${\bar k} \ne 0$ and $r_0^2={\bar k}(n-d-1)(n-2)/(2\Lambda)$ is the solution if $g_{AB}$ is the solution of the $d$-dimensional vacuum Einstein equation with a cosmological constant $\lambda \equiv (d-2)\Lambda/(n-2)$.
\end{The}
In the presence of a cosmological constant, $r_0^2>0$ requires ${\bar k}\Lambda>0$.
These theorems imply that extra-dimensions can be easily attached to the lower-dimensional solutions in order to construct higher-dimensional Kaluza-Klein spacetimes in general relativity.
For example, ${\ma M}^d$ can be the $d$-dimensional Kerr and Kerr-(A)dS spacetimes for $\Lambda=0$ and $\Lambda \ne 0$, respectively.

Next we consider the case in Einstein-Gauss-Bonnet gravity ($\alpha \ne 0$).
In this case, the constraint ${\ma G}^a_{~~b}=0$ is more restrictive than in general relativity.
${\ma G}^A_{~~B}=0$ has the similar structure as the vacuum Einstein equation with a cosmological constant except for the term $\overset{(d)}{H}{}^A_{~B}$, which vanishes identically for $d \le 4$.
Hereafter we consider the case with $d = 3,4$. 
If $r_0^2 \ne -2{\bar k}\alpha(n-d)(n-d-1)$, then Eqs.~(\ref{dec1}) and (\ref{dec2}) demand the Kretschmann invariant constructed from $g_{AB}$ to be constant. 
Thus, there can be no mass point and consequently there is no scalar polynomial singularity on ${\ma M}^d$. 
This is an important implication of the scalar constraint given by the extra dimensional space.

On the other hand, the case with $r_0^2=-2{\bar k}\alpha(n-d)(n-d-1)$ is exceptional which gives the new black hole solution. 
We can generalize the no-go theorem for $d=4$ and $\Theta=0$ established in~\cite{md2006} into the case with $d=3,4$ and non-zero $\Theta$:
\begin{The} 
\label{the:3}
For $d=3$ or $4$, if (i) $r_0^2=-2{\bar k}\alpha(n-d)(n-d-1)$ and (ii) $\alpha\Lambda = -[n^2-(2d-3)n+(d^2-3d-6)]/[8(n-d)(n-d-1)]+\Theta/[8(n-d)(n-d-1)^2]$, then  ${\ma G}^A_{~~B} = 0$ for $n \ge d+2$ and ${\bar k}$ being non-zero.
\end{The}
The proof simply follows from substitution of the conditions (i) and (ii) in  Eq.~(\ref{dec1}). 
With conditions (i) and (ii), which imply ${\bar k} = -1$ and $\Lambda < 0$ for $\alpha > 0$ (for $\alpha < 0$, the opposite will be true with ${\bar k} = 1$ and $\Lambda > 0$), the governing equation for vacuum is a single scalar equation on ${\ma M}^d$, ${\ma G}^{a}_{~~b} = 0$, which is given by
\begin{eqnarray}
0&=&\frac{1}{n-d}\overset{(d)}{R}+\frac{\alpha}{2} \overset{(d)}{L}_{GB}+\frac{2n-3-2d}{\alpha(n-d)^2(n-d-1)} \nonumber \\
&&-\frac{\Theta}{2\alpha (n-d)^2(n-d-1)^2}. \label{KKbasic}
\end{eqnarray}
Thus, solutions with local topology ${\ma M}^d \times {\ma K}^{n-d}$ corresponding to ${\bar k}=-1$ are obtained by solving the equation (\ref{KKbasic}) for $g_{AB}$.

\section{Exact solutions}
\subsection{Schwarzschild-like solution}
With conditions (i) and (ii) in Theorem~\ref{the:3}, the metric on ${\ma M}^d$ is determined only by Eq.~(\ref{KKbasic}) for $d=3,4$.
The warp factor $r_0^2$ is proportional to the Gauss-Bonnet coupling constant $\alpha$ which is supposed to be of the order of the square of the Planck length. 
Thus, compactifying ${\ma K}^{n-d}$ by appropriate identifications, we obtain the Kaluza-Klein spacetime with small and compact extra dimensions.

We seek a static solution for $d=4$ with the metric on ${\ma M}^4$ reading as:
\begin{equation}
g_{AB}dx^Adx^B=-f(r)dt^2+\frac{1}{f(r)}dr^2+r^2d\Sigma_{2(k)}^2, \label{NdS} 
\end{equation}
where $d\Sigma_{2(k)}^2$ is the line element on the unit Einstein space ${\ma K}^2$ and $k= \pm 1, 0$.
Then, Eq.~(\ref{KKbasic}) yields the general solution for the function $f(r)$: 
\begin{widetext}
\begin{eqnarray}
f(r)&\equiv&k+\frac{r^2}{2(n-4)\alpha}\biggl[1\mp\biggl\{1-\frac{2(n-5)(2n-11)-\Theta}{6(n-5)^2}+\frac{4(n-4)^2\alpha^{3/2}M}{r^3}-\frac{4(n-4)^2\alpha^2q}{r^4}\biggl\}^{1/2}\biggl], \label{special}
\end{eqnarray}
\end{widetext}
where $M$ and $q$ are arbitrary dimensionless constants.
Throughout this paper, we always normalize integration constants by $\alpha$.
This is a generalization of the solution obtained in~\cite{md2006} with $\Theta=0$. 

The solution does not have the general relativistic limit $\alpha \to 0$.
There are two branches of the solution corresponding to the sign in front of the square root in Eq.~(\ref{special}), which we call the minus- and plus-branches. 
The metric on ${\ma M}^4$ is asymptotically Reissner-Nordstr\"om-(A)dS spacetime for $k=1$ in spite of absence of Maxwell field since the function $f(r)$ is expanded for $r \to \infty$ as
\begin{widetext}
\begin{eqnarray}
f(r)\approx \frac{r^2}{2(n-4)\alpha}\left(1\mp\sqrt{\frac{2(n-4)(n-5)+\Theta}{6(n-5)^2}}\right)+k\mp \sqrt{\frac{6(n-4)^2(n-5)^2}{2(n-4)(n-5)+\Theta}}\biggl(\frac{\alpha^{1/2} M}{r}- \frac{\alpha q}{r^2}\biggl). \label{infty}
\end{eqnarray}
\end{widetext}
 
For $\Theta=0$, ${\ma M}^4$ is asymptotically AdS.
On the other hand, we can obtain asymptotically flat or de~Sitter spacetime in the minus-branch for $n \ge 8$ if
\begin{eqnarray}
\Theta\ge 2(2n-11)(n-5) \label{flatcond}
\end{eqnarray}
holds with equality corresponding to the asymptotically flat case.
How is this desirable result obtained?
The Weyl term $\Theta$ appears only in $R_{\mu}^{~\alpha\beta\gamma}R_{\nu\alpha\beta\gamma}$ and its contraction terms in Eq.~(\ref{def-H}).
As a result, $\Theta$ acts as a cosmological constant but its value is different on ${\ma M}^4$ and ${\ma K}^{n-4}$, as can be seen in Eqs.~(\ref{dec1}) and (\ref{dec2}).
Consequently, we can choose the value of $\Theta$ such that the effective cosmological constant in Eq.~(\ref{KKbasic}) becomes zero or positive.
For the generalized Bohm metric~(\ref{bohm-h}), the inequality (\ref{flatcond}) is satisfied only for $5 \le n \le 7$ while non-zero $\Theta$ requires $n \ge 8$, and hence there cannot exist a compatible $n$ giving rise to asymptotically flat or dS on ${\ma M}^4$ in this case.

Eq.~(\ref{infty}) suggests that $M$ corresponds to the mass of the central object.
On the other hand, $q$ corresponds to the square of the charge, however it can be both positive and negative.
We call $q$ the ``gravitational charge'', which is similar to Weyl charge of black hole on brane~\cite{dmpr} and it is generated by our choice of the topology of spacetime, splitting it into a product of the usual four-dimensional spacetime and an Einstein space. 
One of the desirable features of Einstein-Gauss-Bonnet gravity for $n=5$, as demonstrated by Boulware-Deser-Wheeler solution~\cite{BDW}, is regularity of metric and weakening of singularity. 
It is interesting that this solution brings this feature down to $\ma M^{4}$. 
In particular note that $f(0) = 1\mp\sqrt{-q}$ for $k=1$, which produces a solid angle deficit and it represents a spacetime of global monopole~\cite{bv}. 
This means that at $r = 0$ curvatures will diverge only as $1/r^2$ and so would be density which on integration over volume will go as $r$ and would therefore vanish. This indicates that singularity is weak as curvatures do not diverge strongly enough. 
Clearly the global structure of this solution is very rich and complex and will be discussed in detail elsewhere~\cite{next}.

Now we explain the origin of the gravitational charge $q$.
For the metric in the form of Eq.~(\ref{NdS}), one just requires one second-order differential equation (\ref{KKbasic}) to determine the metric fully and it will in general have two constants of integration. 
On the other hand, the trace of the Einstein-Gauss-Bonnet equation (\ref{beq}) is given by 
\begin{equation}
-\frac{n-2}{2}R-\frac{(n-4)\alpha}{2}L_{GB}+n\Lambda=\kappa_n^2 T. \label{beq-t}
\end{equation}
The basic equation (\ref{KKbasic}) for $g_{AB}$ resembles this equation with $T=0$ and $\Lambda=\Lambda_{\rm eff}$ defined by 
\begin{eqnarray}
\Lambda_{\rm eff} \equiv \frac{C[\Theta-2(n-d-1)(2n-3-2d)]}{2\alpha (n-d)^2(n-d-1)^2}, \label{Lambdaeff}
\end{eqnarray}
where $C$ is some positive constant.
Thus, Eq.~(\ref{KKbasic}) will generate a Maxwell-like charge because vanishing trace is characteristic of electromagnetic field in four dimensions, i.e., $d=4$. 
This happens only in $d=4$ because electromagnetic stress tensor is not trace-free in other dimensions. 
That is why it is not surprising  that there occurs Maxwell-like additional gravitational charge in solution (\ref{special}). Note that Maxwell-like Weyl charge for black hole on the brane is also caused by vanishing trace~\cite{dmpr}. 

Furthermore, $d=4$ is also crucial for bringing Gauss-Bonnet effects down to four dimensions. 
The second term in the right-hand-side of Eq.~(\ref{KKbasic}) is Gauss-Bonnet term.
Eq.~(\ref{def-H}) gives
\begin{equation}
H^\mu_{~~\mu}=-\frac{n-4}{2}L_{GB}.
\end{equation}
Since $H^\mu_{~~\nu} \equiv 0$ for $n \le 4$ and hence $L_{GB}$ vanishes identically for $n \le 3$ while it will be non-zero in general for $n = 4$. 
Consequently, the effect of Gauss-Bonnet term appears explicitly on ${\ma M}^d$ only for $d=4$.

Now we give a new exact solution for $d=3$ as a concrete example exhibiting these properties mentioned above.
In this case, the effect of Gauss-Bonnet term does not appear on ${\ma M}^d$ and Eq.~(\ref{KKbasic}) reduces to
\begin{equation}
0=\overset{(3)}{R}-\frac{\Theta-2(n-4)(2n-9)}{2\alpha(n-3)(n-4)^2}, \label{KKbasic3}
\end{equation}
which is similar to the vacuum Einstein equation with $\Lambda$ redefined. 
This gives a solution on ${\ma M}^3$ such that 
\begin{eqnarray}
g_{AB}dx^Adx^B&=&-h(r)dt^2+\frac{dr^2}{h(r)}+r^2d\theta^2, \label{NdS3}\\ 
h(r)&\equiv&-M-\frac{\alpha^{1/2}q}{r} \nonumber \\
&&+\frac{2(n-4)(2n-9)-\Theta}{12\alpha(n-3)(n-4)^2}r^2, \label{special3} 
\end{eqnarray}
where $M$ and $q$ are dimensionless constants. 
In this case, the gravitational charge $q$ is not Maxwell-like, because Maxwell stress tensor is not trace-free in three dimensions and so the solution (\ref{special3}) is not like charged BTZ solution~\cite{btz}.

Even for $d=4$, there is no effect of Gauss-Bonnet term on ${\ma M}^4$ when it is a product manifold such as ${\ma M}^3 \times {\ma R}$ because the governing equation~(\ref{KKbasic}) then reduces to the equation on ${\ma M}^3$ similar to Eq.~(\ref{KKbasic3}). 
One example of such solutions is the black string solution, of which metric on ${\ma M}^4$ is obtained by 
\begin{eqnarray}
g_{AB}dx^Adx^B&=&-g(r)dt^2+\frac{dr^2}{g(r)}+r^2d\theta^2+dz^2, \label{NdS4}\\ 
g(r)&\equiv&-M-\frac{\alpha^{1/2}q}{r} \nonumber \\
&&~~+\frac{2(n-5)(2n-11)-\Theta}{12\alpha(n-4)(n-5)^2}r^2, \label{special4} 
\end{eqnarray}
where $M$ and $q$ are dimensionless constants. 
Even when there is null Gauss-Bonnet contribution, additional gravitational charge $q$ is however always there so long as there is split up of spacetime into a product two spacetimes.

\subsection{Vaidya-like solution}
It is well known that Schwarzschild spacetime could be made to radiate null (Vaidya) radiation by transforming the metric into retarded/advanced time coordinate and then making mass parameter function of the time coordinate. 
It is interesting to note that the same procedure also works here. 
This solution~(\ref{special}) can thus be generalized to include Vaidya radiation and it would be given by
\begin{widetext}
\begin{eqnarray}
g_{AB}dx^Adx^B&=&-{\tilde f}(v,r)dv^2+2dvdr+r^2d\Sigma_{2(k)}^2, \label{NdS-d} \\
{\tilde f}(v,r)&\equiv&k+\frac{r^2}{2(n-4)\alpha}\biggl[1\mp\biggl\{1-\frac{2(n-5)(2n-11)-\Theta}{6(n-5)^2} \nonumber \\
&&~~~~~~~~~+\frac{4(n-4)^2\alpha^{3/2}{\tilde M}(v)}{r^3}-\frac{4(n-4)^2\alpha^2{\tilde q}(v)}{r^4}\biggl\}^{1/2}\biggl], \label{special-d}
\end{eqnarray}
\end{widetext}
where ${\tilde M}(v)$ and ${\tilde q}(v)$ are arbitrary functions.
As expected, this solution is quite similar to the null dust solution with the topology of ${\ma M}^2 \times {\ma K}^{n-2}$~\cite{GBnulldust}.
It is noted that the solutions~(\ref{NdS3}) and (\ref{NdS4}) can also be generalized into the Vaidya-like solution.
All this works because trace of null dust stress tensor vanishes and these solutions follow from solving the vanishing trace equation (\ref{KKbasic}).

\begin{figure}[tbp]
\includegraphics[width=.70\linewidth]{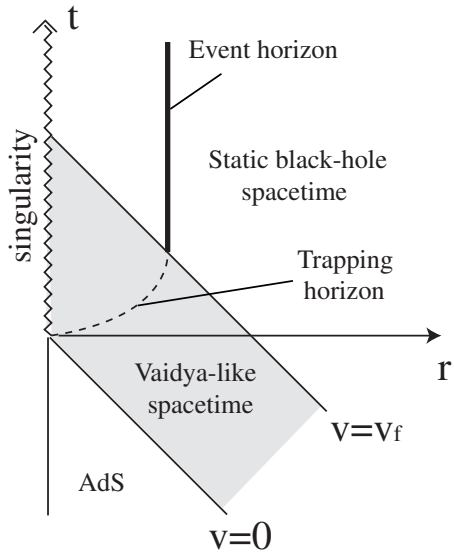}
\caption{
A schematic diagram of the black-hole formation from the AdS spacetime without any matter field.
Here the AdS spacetime for $v<0$ is joined to the static black-hole spacetime (\ref{special}) for $v>v_{\rm f}$ by way of the Vaidya-like spacetime (\ref{NdS-d}).
We set $\Theta=0$, $M(v)=m_0v$ and $q(v)=-1$, where $m_0$ is a positive constant.
A singularity is formed at $v=r=0$ and develops.
}
\label{Fig1}
\end{figure}

This solution manifests gravitational creation of an ingoing charged null dust as another complete example of ``matter without matter''.
Using this solution and the solution~(\ref{special}), we can construct completely vacuum spacetime representing the formation of a black hole from an AdS spacetime by gravitational collapse of a gravitationally created charged null dust (See Fig.~\ref{Fig1}).

\subsection{Taub-NUT-like solution}
There is yet another generalization of our solution (\ref{special}) into a Taub-NUT-like static spacetime on ${\ma M}^4$:
\begin{widetext}
\begin{eqnarray}
g_{AB}dx^Adx^B&=&-F(r)(dt+2l\cos\theta d\phi)^2+\frac{1}{F(r)}dr^2+(r^2+l^2)(d\theta^2+\sin^2\theta d\phi^2), \label{NdS2} \\
F(r)&\equiv&\frac{(r^2+l^2)[r^2+l^2+2(n-4)\alpha]\mp 2\sqrt{(n-4)^2\alpha^{3/2}(r^2+l^2)(r^2-3l^2)(Mr-\alpha^{1/2}q)+A(r)}}{2(n-4)\alpha(r^2-3l^2)}, \label{TNUT} \\
A(r)&\equiv&(n-4)(r^2+l^2)\biggl[4(n-4)\alpha^2l^2+\alpha l^2(5r^2+l^2)+\frac{2(n-4)(n-5)+\Theta}{24(n-4)(n-5)^2}r^6 \nonumber \\
&&~~~~~~~~~~~~+\frac{2(n-4)(n-5)+\Theta}{8(n-4)(n-5)^2}l^2r^4+\frac{3(n-5)(5n-27)-3\Theta}{4(n-4)(n-5)^2}l^4r^2+\frac{1}{4(n-4)}l^6\biggl],
\end{eqnarray}
\end{widetext}
where $l$ is the NUT parameter having the dimension of length and $M$ and $q$ are arbitrary dimensionless constants. As pointed out in~\cite{md2006}, the solution (\ref{special}) bears resemblance with the corresponding Einstein-Maxell-Gauss-Bonnet solution in $n \ge 5$~\cite{GBBH,BDW} and so does the above solution with the corresponding $n(\ge 6)$-dimensional Taub-NUT-like solution in Einstein-Gauss-Bonnet gravity with Maxwell field~\cite{dh2006}.
In the limit of $l \to 0$, we recover the solution (\ref{special}) with $k=1$:
\begin{widetext}
\begin{eqnarray}
F(r)=1+\frac{r^2}{2(n-4)\alpha}\biggl[1\mp\biggl\{1-\frac{2(n-5)(2n-11)-\Theta}{6(n-5)^2}+\frac{4(n-4)^2\alpha^{3/2} M}{r^3}-\frac{4(n-4)^2\alpha^2q}{r^4}\biggl\}^{1/2}\biggl],
\end{eqnarray}
\end{widetext}
so that $M$ and $q$ correspond to mass and gravitational charge, respectively.

\section{General case of Lovelock gravity}
\label{sec:Lovelock}
We consider Einstein-Gauss-Bonnet gravity in this paper.
This theory of gravity is achieved in the low-energy limit of the heterotic superstring theory~\cite{Gross}. The fundamental principle of relativistic theory of gravitation is that its action should be an invariant functional of Riemann curvature. At the same time it should give rise to quasi-linear second order equation of motion. This uniquely identifies Lovelock polynomial of which quadratic term is Gauss-Bonnet term~\cite{lovelock}. For $n\le4$, the linear Einstein Hilbert term suffices as higher order terms make no contribution to equation of motion while for $n\ge5$, higher order terms do contribute and hence they should be included in full description of gravitational dynamics. The order of term to be included in gravitational action thus depends upon dimension of spacetime being considered.
Then, a natural question arises, whether similar Kaluza-Klein solutions exist in Lovelock gravity?

The action for general Lovelock gravity is given by 
\begin{equation} 
\label{actionL}
S=\frac{1}{2\kappa_n^2}\int d^nx\sqrt{-g}\sum_{i=0}\alpha_i{\ma L}_{(i)}+S_{\rm matter},
\end{equation}
where ${\ma L}_{(i)}$ is the $i$-th order Lovelock Lagrangian density which is a polynomial in Riemann curvature and its contractions, and we identify ${\ma L}_{(0)} \equiv -2$, ${\ma L}_{(1)} \equiv R$, ${\ma L}_{(2)} \equiv L_{GB}$ and so on~\cite{lovelock}.
$\alpha_i$ are coupling constants such as $\alpha_0 \equiv \Lambda$, $\alpha_1 \equiv 1$, and $\alpha_2 \equiv \alpha$.
The gravitational equation following from this action is given by 
\begin{equation}
{\ma G}_{\mu\nu} \equiv \sum_{i=0}\alpha_{i}{G}^{(i)}_{\mu\nu}=\kappa_n^2 {T}_{\mu\nu}, \label{beqL}
\end{equation}
where tensors ${G}^{(i)}_{\mu\nu}$ are given from ${\ma L}_{(i)}$ such as ${G}^{(0)}_{\mu\nu} \equiv g_{\mu\nu}$, ${G}^{(1)}_{\mu\nu} \equiv G_{\mu\nu}$, and ${G}^{(i)}_{\mu\nu} \equiv H_{\mu\nu}$.
The two important idetities for the Lovelock tensors $G_{\mu\nu}^{(i)}$ are 
\begin{eqnarray}
G_{\mu\nu}^{(i)}\equiv 0~~\mbox{for}~~n \le 2i \label{pA}
\end{eqnarray}
and
\begin{eqnarray}
{\ma L}_{(i)} \equiv 0~~\mbox{for}~~n \le 2i-1. \label{pB}
\end{eqnarray}
The identity~(\ref{pB}) is proven by the identity~(\ref{pA}) and the formula~\cite{cai}
\begin{eqnarray}
G^{(i)\mu}_{~~~~\mu}=-\frac{n-2i}{2}{\ma L}_{(i)}.
\end{eqnarray}

When we consider the $n$-dimensional Kaluza-Klein vacuum spacetime ${\ma M}^{d} \times {\ma K}^{n-d}$ with the metric $g_{\mu\nu}=\mbox{diag}(g_{AB},r_0^2\gamma_{ab})$, ${\ma G}_{\mu\nu}$ should get decomposed as follows:
\begin{eqnarray}
{\ma G}_{AB}&=&\sum_{i=0}f_{i}(n,d,r_0^2,{\bar k}, \alpha_j, \Theta)\overset{(d)}{G}{}^{(i)}_{AB},\label{dec1L} \\
{\ma G}_{ab}&=&\delta_{ab}\sum_{i=0}g_{i}(n,d,r_0^2,{\bar k}, \alpha_j, \Theta)\overset{(d)}{\ma L}{}_{(i)},\label{dec2L}
\end{eqnarray}
where $f_{(i)}$ and $g_{(i)}$ are functions of their arguments.  
Let us consider the most important case $d=4$.
In this case, the vacuum equation ${\ma G}_{AB}=0$ is similar in structure with Eq.~(\ref{dec1}) with $d=4$ because of the identity~(\ref{pA}).
Thus it would be possible to set $f_{(0)}=f_{(1)} \equiv 0$ in order to make ${\ma G}_{AB} \equiv 0$ vacuous keeping $g_{AB}$ completely free and arbitrary by choosing constants $r_0^2$ and $\alpha_{j}$.
Then, the governing equation is only (\ref{dec2L}), which is similar in structure to Eq.~(\ref{KKbasic}) with $d=4$ because of the identity~(\ref{pB}), i.e., the difference is only in the coefficients in the equation.
From the discussions above, it is clear that structure of the solutions obtained and discussed here is quite generic and similar solutions should therefore exist in general Lovelock gravity.

\section{discussions and conclusion}
In this paper, we obtained new exact solutions in Einstein-Gauss-Bonnet gravity which offer direct and purely classical examples of curvature manifesting as matter, i.e., ``matter without matter''.
The origins of the Maxwell field and a null dust fluid have been proposed.
Considering the Kaluza-Klein split up of the $n$-dimensional spacetime as $\ma M^{d} \times \ma K^{n-d}$, we showed that there exist similar solutions with the recently discovered new black hole solution by the authors~\cite{md2006} only for $d=3,4$.
Further, it was shown that the $d=4$ case is special in the sense that the gravitational creation the Maxwell field as well as Gauss-Bonnet term effect occur explicitly on ${\ma M}^d$ only in this case.

Kaluza-Klein split of spacetime does two things. 
One, it decomposes the equation, Eq~(\ref{beq}) into Eq~(\ref{dec1}) and Eq~(\ref{dec2}) in which the former is the usual Einstein equation with $\Lambda$ redefined while the latter acts as a constraint on it. 
Then it is possible to make Eq~(\ref{dec1}) vacuous by proper prescription for the parameters, $r_0$ and $\Lambda$ in terms of $\alpha$ and so that the metric is fully determined by the constraint, Eq~(\ref{dec2}). 
It was shown that no interior solutions including a matter field with constant $r_0$ can be attached to the solution~(\ref{special})~\cite{md2006}.
Therefore, the warp factor $r_0$ in the static interior solution must depend on $r$. We have also argued in section~\ref{sec:Lovelock} that the structure of our solutions is quite generic in general Lovelock gravity. Thus similar solutions would also exist when other higher order terms are included in Lovelock Lagrangian.

Second, it also facilitates bringing down Gauss-Bonnet effects to four dimensions. 
In general, Gauss-Bonnet term does not contribute anything in four-dimensional spacetime ($n=4$), while it does contribute non-trivially in its analogue in Eq~(\ref{KKbasic}) for $d=4$. 
This is the determining equation for new black holes~(\ref{special}). 
It is Gauss-Bonnet contribution which makes the black hole metric finite and regular everywhere and it weakens the singularity. 
The additional gravitational charge is however caused by Kaluza-Klein split. 

We have successfully generalized the solution~(\ref{special}) into Vaidya-like metric on ${\ma M}^4$.
That solution manifests gravitational creation of an ingoing charged null dust and is another complete example of ``matter without matter''.
Using this solution and the solution~(\ref{special}), we can construct the vacuum spacetime representing the formation of a black hole from the AdS spacetime on ${\ma M}^4$ by the gravitational collapse of a gravitationally created charged null dust.

Also, we have generalized the solution~(\ref{special}) into Taub-NUT-like metric on ${\ma M}^4$.
It may be asked how does the constraint equation permit presence of NUT parameter $l$ which is interpreted as gravitational magnetic charge (see for example~\cite{lz}). 
This happens because it is also the part of spacetime symmetry and there is no stress tensor $T_{AB}$ which generates NUT parameter. 
Similar will be the case for Kerr rotational parameter if we find a corresponding solution for Kerr geometry.
This is an interesting open problem which is under investigation. 
We do however believe that there should exist a rotating analogue of our solution~(\ref{special}). 
The effects of Gauss-Bonnet term on four-dimensional black hole are particularly interesting in relation to singularity, horizon and causal structure.

The global structure of the solution~(\ref{special}) depending on the parameters will be shown in the forthcoming paper~\cite{next}.
In the case that ${\ma K}^{n-4}$ is a space of negative constant curvature, the metric on ${\ma M}^4$ of our solution is asymptotically AdS for $\alpha >0$.
If ${\ma K}^{n-4}$ is an Einstein space of negative curvature satisfying the Einstein-space condition~(\ref{hc}), it is possible to have asymptotically flat or de~Sitter spacetime. However we have not been able to find an explicit example of it. 
Such a solution will represent an isolated black hole or a black hole in the de~Sitter universe.
The generalized Bohm metric~(\ref{bohm-h}) satisfies the Einstein-space condition but does not meet compatibly the requirement for ${\ma M}^4$ to be asymptotically flat or de~Sitter.

In the original Kaluza-Klein theory, the origin of the Maxwell field is the extra-dimensional component of the five-dimensional metric with which the five-dimensional vacuum Einstein equation is decomposed into the four-dimensional Einstein-Maxwell equation~\cite{KK,ow1997}. Here on the other hand, we have given completely different and novel generation of Maxwell field as well as of null dust fluid in the framework of Einstein-Gauss-Bonnet gravity.
This is a partial success to explain origin of all matter in our four-dimensional universe. Since our mechanism works only for trace free matter fields, creation of other matter, especially with non-zero trace remains a very important open problem. Undoubtedly its solution will have a great bearing on our understanding of spacetime and matter.

\acknowledgments
It appears that the results of theorems 1 and 2 are more or less known as folklore, though may not be stated so explicitly, and we thank Hideki Ishihara, Panagiota Kanti, Umpei Miyamoto, and Masato Nozawa for bringing this fact to our notice and also for useful comments. 
HM would like to thank Rong-Gen Cai for useful comments.
The authors would like to thank Albert Einstein Institute, Golm warmly for hospitality which facilitated this work.  
HM also would like to thank Department of Physics, National Central University for warm hospitality.  
HM was supported by the Grant-in-Aid for Scientific Research Fund of the Ministry of Education, Culture, Sports, Science and Technology, Japan (Young Scientists (B) 18740162).

\appendix
\section{generalized Bohm metric and the Einstein-space condition}
We start with the following ansatz for metrics:
\begin{eqnarray}
ds^2=d\rho^2+a(\rho)^2\gamma_{ij}dx^idx^k+b(\rho)^2{\bar \gamma}_{\alpha\beta}dx^\alpha dx^\beta, \label{bohm2}
\end{eqnarray}  
where $\gamma_{ij}$ and ${\bar \gamma}_{\alpha\beta}$ are the unit metric on the $p(\ge 2)$-dimensional and $q(\ge 2)$-dimensional spaces of constant curvature with their curvature $k_1$ and $k_2$, respectively.
The components of the Ricci tensor are given by
\begin{eqnarray}
R_{11}&=&-p\frac{{\ddot a}}{a}-q\frac{{\ddot b}}{b}, \\
R_{ij}&=&a^2\gamma_{ij}\left[-\frac{{\ddot a}}{a}-q\frac{{\dot a}}{a}\frac{{\dot b}}{b}+\frac{(p-1)(k_1-{\dot a}^2)}{a^2}\right], \\
R_{\alpha\beta}&=&b^2{\bar \gamma}_{\alpha\beta}\left[-\frac{{\ddot b}}{b}-p\frac{{\dot a}}{a}\frac{{\dot b}}{b}+\frac{(q-1)(k_2-{\dot b}^2)}{b^2}\right], 
\end{eqnarray}  
where $x^1 \equiv \rho$ and a dot denotes the derivative with respect to $\rho$.
The Einstein-${\tilde \Lambda}$ equation $R_{\mu\nu}={\tilde \Lambda}g_{\mu\nu}$ give rise to two second-order differential equations for $a $ and $b$
\begin{eqnarray}
{\tilde \Lambda}&=&-\frac{{\ddot a}}{a}-q\frac{{\dot a}}{a}\frac{{\dot b}}{b}+\frac{(p-1)(k_1-{\dot a}^2)}{a^2}, \label{b1}\\
{\tilde \Lambda}&=&-\frac{{\ddot b}}{b}-p\frac{{\dot a}}{a}\frac{{\dot b}}{b}+\frac{(q-1)(k_2-{\dot b}^2)}{b^2} \label{b2}
\end{eqnarray}  
together with the first-order constraint
\begin{eqnarray}
&&2pq\frac{{\dot a}}{a}\frac{{\dot b}}{b}-\frac{p(p-1)(k_1-{\dot a}^2)}{a^2}-\frac{q(q-1)(k_2-{\dot b}^2)}{b^2} \nonumber \\
&&~~~~~~~=-(p+q-1){\tilde \Lambda}. \label{b3}
\end{eqnarray}  
We assume $b=b_0$, where $b_0$ is a constant.
Then, Eq.~(\ref{b2}) gives
\begin{eqnarray}
b_0&=&\sqrt{\frac{(q-1)k_2}{{\tilde \Lambda}}}. \label{b0}
\end{eqnarray}  
With Eq.~(\ref{b0}), Eqs.~(\ref{b1}) and (\ref{b3}) give
\begin{eqnarray}
0&=&\frac{{\ddot a}}{a}+\frac{{\tilde \Lambda}}{p}, \label{b4}\\
{\tilde \Lambda}&=&\frac{p(k_1-{\dot a}^2)}{a^2}. \label{b5}
\end{eqnarray}  

A Bohm metric Bohm$(p,q)_1$~\cite{bohm1998,ghp2003} is the special solution for $k_1=k_2=1$ and ${\tilde\Lambda>0}$:
\begin{eqnarray}
a(\rho)&=&\sqrt{\frac{p}{\tilde \Lambda}}\sin\biggl(\sqrt{\frac{\tilde\Lambda}{p}}\rho\biggl), \label{bohm-s1} \\
b(\rho)&=&\sqrt{\frac{q-1}{{\tilde \Lambda}}}. \label{bohm-s2}
\end{eqnarray}  

Now we consider the case with $k_2=-1$ and ${\tilde\Lambda<0}$. 
For $k_1=-1$, there exists a counterpart of Bohm$(p,q)_1$ with negative curvature:
\begin{eqnarray}
a(\rho)&=&\sqrt{\frac{p}{-\tilde \Lambda}}\cosh\biggl(\sqrt{\frac{-\tilde\Lambda}{p}}\rho\biggl), \label{bohm-h1} \\
b(\rho)&=&\sqrt{\frac{q-1}{-{\tilde \Lambda}}}. \label{bohm-h2}
\end{eqnarray}  

In these two cases, we obtain
\begin{eqnarray}
C^{1\mu\nu\sigma}C_{1\mu\nu\sigma}&=&2p\frac{{\ddot a}^2}{a^2}+2q\frac{{\ddot b}^2}{b^2}-\frac{2{\tilde\Lambda}^2}{p+q},\\
&=&\frac{2q{\tilde\Lambda}^2}{p(p+q)},\label{cc1} \\
C^{i\mu\nu\sigma}C_{j\mu\nu\sigma}&=&\delta^i_j\biggl[2\frac{{\ddot a}^2}{a^2}+\frac{2(p-1)}{a^4}(k_1-{\dot a}^2)^2 \nonumber \\
&&~~~-\frac{2{\tilde\Lambda}^2}{p+q}\biggl], \\
&=&\frac{2q{\tilde\Lambda}^2}{p(p+q)}\delta^i_j,\label{cc2} \\
C^{\alpha\mu\nu\sigma}C_{\beta\mu\nu\sigma}&=&\delta^i_j\biggl[2\frac{{\ddot b}^2}{b^2}+\frac{2(q-1)}{b^4}(k_2-{\dot b}^2)^2 \nonumber \\
&&~~~-\frac{2{\tilde\Lambda}^2}{p+q}\biggl], \\
&=&\frac{2(p+1){\tilde\Lambda}^2}{(q-1)(p+q)}\delta^\alpha_\beta,\label{cc3} 
\end{eqnarray}
so that the Einstein-space condition (\ref{hc}) holds only when $p=q-1(\ge 3)$ is satisfied.

In the main text, we consider these metrics on the $(n-d)$-dimensional submanifold. 
Thus, we have $n-d=p+q+1(=2q)$ and ${\tilde\Lambda}={\bar k}(p+q)$.
Finally, from Eqs.~(\ref{cc1}), (\ref{cc2}), and (\ref{cc3}), the constant $\Theta$ in Eq.~(\ref{hc}) is obtained by 
\begin{eqnarray}
\Theta=\frac{2q(2q-1){\bar k}^2}{q-1}.
\end{eqnarray}

\section{Tensor decomposition}
We consider the $n$-dimensional spacetime ${\ma M}^n \approx {\ma M}^d\times {\ma K}^{n-d}$ with the general metric 
\begin{eqnarray}
g_{\mu\nu}=\mbox{diag}(g_{AB},r^2\gamma_{ab}),
\label{eq:structure}
\end{eqnarray}
where $g_{AB}$ is an arbitrary Lorentz metric on ${\ma M}^d$, $r$ is a scalar function on ${\ma M}^d$ with $r=0$ defining the boundary of ${\ma M}^d$, and $\gamma_{ab}$ is the unit metric on the $(n-d)$-dimensional Einstein space ${\ma K}^{n-d}$. 
We introduce the covariant derivatives on spacetime ${\ma M}^n$, the subspacetime ${\ma M}^d$ and the Einstein space ${\ma K}^{n-d}$ with
\begin{eqnarray}
g_{\mu\nu;\lambda}&=&0,\\
g_{AB|C}&=&0,\\
g_{ab:c}&=&0.
\end{eqnarray}

Non-zero component of the Christoffel symbol is 
\begin{eqnarray}
\overset{(n)}{\Gamma}{}^A_{BD}&=&\overset{(d)}{\Gamma}{}^A_{BD}, \\
\overset{(n)}{\Gamma}{}^a_{bd}&=&\overset{(n-d)}{\Gamma}{}^a_{bd}, \\
\overset{(n)}{\Gamma}{}^A_{bd}&=&-rr^{|A}\gamma_{bd}, \\
\overset{(n)}{\Gamma}{}^a_{Bd}&=&\frac{r_{|B}}{r}\delta^a_d, 
\end{eqnarray}  
where the superscripts $(n)$, $(d)$, and $(n-d)$ imply the geometrical quantities on ${\ma M}^n$, ${\ma M}^d$, and ${\ma K}^{n-d}$, respectively.
Our definition of the Riemann tensor is
\begin{eqnarray}
R^\mu_{~~\nu\rho\sigma}&=&\Gamma^\mu_{~~\nu\sigma,\rho}-\Gamma^\mu_{~~\nu\rho,\sigma} \nonumber \\
&&+\Gamma^\mu_{~~\kappa\rho}\Gamma^\kappa_{~~\nu\sigma}-\Gamma^\mu_{~~\kappa\sigma}\Gamma^\kappa_{~~\nu\rho}.
\end{eqnarray}
The $(n-d)$-dimensional Einstein space satisfies
\begin{eqnarray}
\overset{(n-d)}{R}{}_{abde}=\overset{(n-d)}{C}{}_{abde}+{\bar k}(\gamma_{ad}\gamma_{be}-\gamma_{ae}\gamma_{bd}),
\end{eqnarray}
which is contracted to give
\begin{eqnarray}
\overset{(n-d)}{R}{}_{ab}&=&{\bar k}(n-d-1)\gamma_{ab},\\
\overset{(n-d)}{R}&=&{\bar k}(n-d)(n-d-1),
\end{eqnarray}  
where $\overset{(n-d)}{C}_{abde}$ is the Weyl tensor and ${\bar k}=\pm 1,0$.

Non-zero component of the Riemann tensor is
\begin{eqnarray}
\overset{(n)}{R}{}^A_{~~BDE}&=&\overset{(d)}{R}{}^A_{~~BDE},\\
\overset{(n)}{R}{}^a_{~~bde}&=&({\bar k}-r_{|A}r^{|A})(\delta^a_d\gamma_{be}-\delta^a_e\gamma_{bd}) \nonumber \\
&&+\overset{(n-d)}{C}{}^a_{~~bde},\\
\overset{(n)}{R}{}^A_{~~bDe}&=&-rr^{|A}_{~~|D}\gamma_{be},\\
\overset{(n)}{R}{}^a_{~~BDe}&=&\frac{r_{|B|D}}{r}\delta^a_e.
\end{eqnarray}
Non-zero component of the Ricci tensor is 
\begin{eqnarray}
\overset{(n)}{R}{}_{BE}&=&\overset{(d)}{R}{}_{BE}-(n-d)\frac{r_{|B|E}}{r},\\
\overset{(n)}{R}{}_{be}&=&\gamma_{be}\biggl[-rr^{|A}_{~~|A} \nonumber \\
&&~~~~~+(n-d-1)({\bar k}-r_{|A}r^{|A})\biggl].
\end{eqnarray}  
The Ricci scalar is 
\begin{eqnarray}
\overset{(n)}{R}&=&\overset{(d)}{R}-2(n-d)\frac{r^{|A}_{~~|A}}{r} \nonumber \\
&&+\frac{(n-d)(n-d-1)}{r^2}({\bar k}-r_{|A}r^{|A}).
\end{eqnarray} 
Non-zero component of the Einstein tensor is
\begin{eqnarray}
\overset{(n)}{G}{}^B_{~~E}&=&\overset{(d)}{G}{}^B_{~E}-(n-d)\frac{r^{|B}_{~~|E}}{r}+\delta^B_E\biggl[(n-d)\frac{r^{|A}_{~~|A}}{r} \nonumber \\
&&-\frac{(n-d)(n-d-1)}{2r^2}({\bar k}-r_{|A}r^{|A})\biggl],\\
\overset{(n)}{G}{}^b_{~~e}&=&\delta^b_e\biggl[-\frac12\overset{(d)}{R}+(n-d-1)\frac{r^{|A}_{~~|A}}{r} \nonumber \\
&&-\frac{(n-d-1)(n-d-2)}{2r^2}({\bar k}-r_{|A}r^{|A})\biggl].
\end{eqnarray}

Now we assume the {\it Einstein-space condition} originally introduced in the case of $d=2$ by Dotti and Gleiser~\cite{dg2005,dg}
\begin{eqnarray}
\overset{(n-d)}{C}{}^{abde}\overset{(n-d)}{C}{}_{fbde}&=&\Theta \delta^a_{~~f}, 
\end{eqnarray}  
where $\Theta$ is a scalar on ${\cal K}^{n-d}$.
The identity $\overset{(n)}{H}{}^{a\nu}_{~~;\nu}=0$ gives $\Theta_{,a}=0$, and consequently $\Theta$ is a dimensionless constant.
The Weyl tensor vanishes identically in three or less dimensions, so that $\Theta \equiv 0$ holds for $n \le d+3$.

For later convenience, we obtain
\begin{widetext}
\begin{eqnarray}
\overset{(n)}{R}{}^{B\mu}\overset{(n)}{R}{}_{\mu E}&=&\biggl[\overset{(d)}{R}{}^{BD}-(n-d)\frac{r^{|B|D}}{r}\biggl]\biggl[\overset{(d)}{R}{}_{DE}-(n-d)\frac{r_{|D|E}}{r}\biggl],\\
\overset{(n)}{R}{}^{B\mu}\overset{(n)}{R}{}_{\mu e}&=&\overset{(n)}{R}{}^{b\mu}\overset{(n)}{R}{}_{\mu E}=0,\\
\overset{(n)}{R}{}^{b\mu}\overset{(n)}{R}{}_{\mu e}&=&\frac{1}{r^4}\delta^b_e\biggl[-rr^{|A}_{~~|A}+(n-d-1)({\bar k}-r_{|A}r^{|A})\biggl]^2,
\end{eqnarray}
\begin{eqnarray}
\overset{(n)}{R}{}^{\mu\nu}\overset{(n)}{R}{}^A_{~~\mu D \nu}&=&\biggl(\overset{(d)}{R}{}^{BE}-(n-d)\frac{r^{|B|E}}{r}\biggl)\overset{(d)}{R}{}^{A}_{~~BDE}-\frac{n-d}{r^3}r^{|A}_{~~|D}\biggl(-rr^{|E}_{~~|E}+(n-d-1)({\bar k}-r_{|E}r^{|E})\biggl),\\
\overset{(n)}{R}{}^{\mu\nu}\overset{(n)}{R}{}^a_{~~\mu d \nu}&=&\delta^a_d\biggl[-\frac{r_{|B|E}}{r}\biggl(\overset{(d)}{R}{}^{BE}-(n-d)\frac{r^{|B|E}}{r}\biggl) \nonumber \\
&&~~~~~~~~~~~+\frac{n-d-1}{r^4}({\bar k}-r_{|B}r^{|B})\{-rr^{|A}_{~~|A}+(n-d-1)({\bar k}-r_{|A}r^{|A})\}\biggl],
\end{eqnarray}
\begin{eqnarray}
\overset{(n)}{R}{}^{A\mu\nu\rho}\overset{(n)}{R}{}_{B\mu\nu\rho}&=&\overset{(d)}{R}{}^{ADEF}\overset{(d)}{R}{}_{BDEF}+\frac{2(n-d)}{r^2}r^{|A|E}r_{|B|E},\\
\overset{(n)}{R}{}^{A\mu\nu\rho}\overset{(n)}{R}{}_{b\mu\nu\rho}&=&\overset{(n)}{R}{}^{a\mu\nu\rho}\overset{(n)}{R}{}_{B\mu\nu\rho}=0,\\
\overset{(n)}{R}{}^{a\mu\nu\rho}\overset{(n)}{R}{}_{b\mu\nu\rho}&=&\delta^a_b\biggl[\frac{2}{r^2}r^{|D|E}r_{|D|E}+\frac{2(n-d-1)}{r^4}({\bar k}-r^{|A}r_{|A})^2+\frac{\Theta}{r^4}\biggl],
\end{eqnarray}
\begin{eqnarray}
\overset{(n)}{L}_{GB}&=&\overset{(d)}{L}_{GB}+(n-d)\biggl[2\overset{(d)}{R}-2(n-d)\frac{r^{|A}_{~~|A}}{r}+\frac{(n-d)(n-d-1)}{r^2}({\bar k}-r_{|A}r^{|A})\biggl] \nonumber \\
&&\times \biggl[-2\frac{r^{|D}_{~~|D}}{r}+\frac{(n-d-1)}{r^2}({\bar k}-r_{|D}r^{|D})\biggl]+8(n-d)\overset{(d)}{R}{}_{BD}\frac{r^{|B|D}}{r}  \nonumber \\
&&-\frac{4(n-d)(n-d-1)}{r^2}r_{|B|D}r^{|B|D}-\frac{4(n-d)}{r^4}\biggl[-rr^{|A}_{~~|A}+(n-d-1)({\bar k}-r_{|A}r^{|A})\biggl]^2 \nonumber \\
&&+\frac{2(n-d)(n-d-1)}{r^4}({\bar k}-r_{|A}r^{|A})^2+\frac{(n-d)\Theta}{r^4}.
\end{eqnarray}
\end{widetext}
The Kretschmann invariant is
\begin{eqnarray}
\overset{(n)}{K}&=&\overset{(d)}{K}+\frac{4(n-d)}{r^2}r^{|A|D}r_{|A|D}+\frac{(n-d)\Theta}{r^4} \nonumber \\
&&+\frac{2(n-d)(n-d-1)}{r^4}({\bar k}-r_{|A}r^{|A})^2.
\end{eqnarray}

Using these expressions, we finally obtain
\begin{widetext}
\begin{eqnarray}
\overset{(n)}{H}{}^B_{~~E}&=&2\biggl(\overset{(d)}{R}{}^B_{~~E}-(n-d)\frac{r^{|B}_{~~|E}}{r}\biggl)\biggl[\overset{(d)}{R}-2(n-d)\frac{r^{|A}_{~~|A}}{r}+\frac{(n-d)(n-d-1)}{r^2}({\bar k}-r_{|A}r^{|A})\biggl] \nonumber \\
&&-4\biggl[\overset{(d)}{R}{}^{BD}-(n-d)\frac{r^{|B|D}}{r}\biggl]\biggl[\overset{(d)}{R}{}_{DE}-(n-d)\frac{r_{|D|E}}{r}\biggl] \nonumber \\
&&-4\biggl(\overset{(d)}{R}{}^{AD}-(n-d)\frac{r^{|A|D}}{r}\biggl)\overset{(d)}{R}{}^{B}_{~~AED}+\frac{4(n-d)}{r^3}r^{|B}_{~~|E}\biggl(-rr^{|F}_{~~|F}+(n-d-1)({\bar k}-r_{|A}r^{|A})\biggl) \nonumber \\
&&+2\biggl(\overset{(d)}{R}{}^{BADF}\overset{(d)}{R}{}_{EADF}+\frac{2(n-d)}{r^2}r^{|B|F}r_{|E|F}\biggl)-\frac12 \overset{(n)}{L}_{GB}\delta^B_E,\\
\overset{(n)}{H}{}^B_{~~e}&=&\overset{(n)}{H}{}^b_{~~E}=0,\\
\overset{(n)}{H}{}^b_{~~e}/\delta^b_e&=&\frac{2}{r^2}\biggl[-rr^{|D}_{~~|D}+(n-d-1)({\bar k}-r_{|D}r^{|D})\biggl]\biggl[\overset{(d)}{R}-2(n-d)\frac{r^{|A}_{~~|A}}{r}+\frac{(n-d)(n-d-1)}{r^2}({\bar k}-r_{|A}r^{|A})\biggl] \nonumber \\
&&-\frac{4}{r^4}\biggl[-rr^{|A}_{~~|A}+(n-d-1)({\bar k}-r_{|A}r^{|A})\biggl]^2 \nonumber \\
&&-4\biggl[-\frac{r_{|B|E}}{r}\biggl(\overset{(d)}{R}{}^{BE}-(n-d)\frac{r^{|B|E}}{r}\biggl)+\frac{(n-d-1)}{r^4}({\bar k}-r_{|E}r^{|E})\{-rr^{|A}_{~~|A}+(n-d-1)({\bar k}-r_{|A}r^{|A})\}\biggl] \nonumber \\
&&+2\biggl[\frac{2}{r^2}r^{|D|E}r_{|D|E}+\frac{2(n-d-1)}{r^4}({\bar k}-r^{|A}r_{|A})^2\biggl]+\frac{2\Theta}{r^4}-\frac12 \overset{(n)}{L}_{GB}.
\end{eqnarray}  
\end{widetext}
It is noted that not $\overset{(d)}{L}_{GB}$ but $\overset{(n)}{L}_{GB}$ is included in above equations.

Let us consider the case with $r=r_0$, where $r_0$ is a positive constant.
Then, we obtain
\begin{eqnarray}
\overset{(n)}{G}{}^B_{~~E}&=&\overset{(d)}{G}{}^B_{~E}-\frac{(n-d)(n-d-1){\bar k}}{2r_0^2}\delta^B_E,\\
\overset{(n)}{G}{}^B_{~~e}&=&\overset{(n)}{G}{}^b_{~~E}=0,\\
\overset{(n)}{G}{}^b_{~~e}&=&\delta^b_e\biggl[-\frac12\overset{(d)}{R}-\frac{(n-d-1)(n-d-2){\bar k}}{2r_0^2}\biggl].
\end{eqnarray}
\begin{eqnarray}
\overset{(n)}{L}_{GB}&=&\overset{(d)}{L}_{GB}+\frac{2(n-d)(n-d-1){\bar k}}{r_0^2}\overset{(d)}{R} \nonumber \\
&&+\frac{(n-d)(n-d-1)(n-d-2)(n-d-3){\bar k}^2}{r_0^4} \nonumber \\
&&+\frac{(n-d)\Theta}{r_0^4}.
\end{eqnarray}
\begin{widetext}
\begin{eqnarray}
\overset{(n)}{H}{}^B_{~~E}&=&\overset{(d)}{H}{}^B_{~~E}+\frac{2(n-d)(n-d-1){\bar k}}{r_0^2}\overset{(d)}{G}{}^B_{~~E} \nonumber \\
&&~~~~~~-\left[\frac{(n-d)(n-d-1)(n-d-2)(n-d-3){\bar k}^2}{2r_0^4}+\frac{n-d}{2r_0^4}\Theta\right]\delta^B_E,\\
\overset{(n)}{H}{}^B_{~~e}&=&\overset{(n)}{H}{}^b_{~~E}=0,\\
\overset{(n)}{H}{}^b_{~~e}/\delta^b_e&=&-\frac{{\bar k}(n-d-1)(n-d-2)}{r_0^2}\overset{(d)}{R}-\frac{(n-d-1)(n-d-2)(n-d-3)(n-d-4){\bar k}^2}{2r_0^4} \nonumber \\
&&~~~~~~-\frac{n-d-4}{2r_0^4}\Theta-\frac12 \overset{(d)}{L}_{GB}.
\end{eqnarray}  
\end{widetext}
The Kretschmann invariant is
\begin{equation}
\overset{(n)}{K}=\overset{(d)}{K}+\frac{2(n-d)(n-d-1){\bar k}^2}{r_0^4}+\frac{(n-d)\Theta}{r_0^4}.
\end{equation}
Thus, we finally obtain
\begin{widetext}
\begin{eqnarray}
{\cal G}^B_{~E}&=&\biggl[1+\frac{2(n-d)(n-d-1)\alpha {\bar k}}{r_0^2}\biggl]\overset{(d)}{G}{}^B_{~E}+\alpha \overset{(d)}{H}{}^B_{~E} \nonumber \\
&&+\biggl[\Lambda-\frac{(n-d)(n-d-1){\bar k}}{2r_0^2}-\frac{(n-d)(n-d-1)(n-d-2)(n-d-3)\alpha {\bar k}^2}{2r_0^4}-\frac{(n-d)\alpha}{2r_0^4}\Theta\biggl]\delta^B_E,\\
{\cal G}^B_{~e}&=&{\cal G}^b_{~E}=0,\\
{\cal G}^b_{~e}/\delta^b_{~e}&=&-\biggl[\frac12+\frac{{\bar k}\alpha(n-d-1)(n-d-2)}{r_0^2}\biggl]\overset{(d)}{R}-\frac{\alpha}{2}\overset{(d)}{L}_{GB}+\Lambda-\frac{(n-d-4)\alpha}{2r_0^4}\Theta \nonumber \\
&&-\frac{(n-d-1)(n-d-2){\bar k}}{2r_0^2}-\frac{\alpha {\bar k}^2(n-d-1)(n-d-2)(n-d-3)(n-d-4)}{2r_0^4}.
\end{eqnarray}
\end{widetext}



\begin{references}
\bibitem{KK} 
   T.~Kaluza, 
   Sitzungsber. Preuss. Akad. Wiss. Berlin (Math. Phys.) {\bf K1}, 966 (1921).
   O.~Klein, 
   Z. Phys. {\bf 37}, 895 (1926).
\bibitem{ow1997} 
J.M.~Overduin and P.S.~Wesson, 
Phys. Rep. {\bf 283}, 303 (1997). 
\bibitem{md2006} 
   H.~Maeda and N.~Dadhich, 
   Phys. Rev. D {\bf 74}, 021501(R) (2006).
\bibitem{mal} 
J.~Maldacena, 
Adv. Theor. Math. Phys. {\bf 2}, 231 (1998).
\bibitem{superstring}
J.~Polchinski, 
{\it String Theory: An Introduction to the Bosonic String}
(Cambridge University Press, Cambridge, England, 1998);
{\it String Theory: Superstring Theory and Beyond}
(Cambridge University Press, Cambridge, England, 1998).
\bibitem{Lukas} 
   P. Ho\v rava and E. Witten, 
   Nucl. Phys. {\bf B475}, 94 (1996);
   A. Lukas, B. A. Ovrut, K.S. Stelle, and D. Waldram,
   Phys. Rev. D {\bf 59}, 086001 (1999);
   A. Lukas, B. A. Ovrut, and D. Waldram,
   Phys. Rev. D {\bf 60}, 086001 (1999).
\bibitem{dad} 
   N.~Dadhich, 
   hep-th/0509126.
\bibitem{lovelock}
D. Lovelock,
J. Math. Phys. {\bf 12} 498 (1971).
\bibitem{Gross} 
D.J.~Gross and J.H.~Sloan, 
Nucl. Phys. {\bf B291}, 41 (1987);\\
M.C.~Bento and O.~Bertolami,
Phys. Lett. {\bf B368}, 198 (1996).
\bibitem{olea2005} 
R.~Olea, 
JHEP {\bf 0506}, 023 (2005).
\bibitem{acotz2000} 
R.~Aros, M.~Contreras, R.~Olea, R.~Troncoso, and J.~Zanelli,
Phys. Rev. Lett. {\bf 84}, 1647 (2000).
\bibitem{Gravitation}
C.W.~Misner, K.S.~Thorne, and J.A.~Wheeler,
{\it ``Gravitation''}, (Freeman, San Francisco, 1973).
\bibitem{dg2005} 
G.~Dotti and R.J.~Gleiser, 
   Phys. Lett. {\bf B627}, 174 (2005).
\bibitem{dg} 
In~\cite{dg2005}, this condition with $d=2$ is called the {\it horizon condition}.
\bibitem{bohm1998} 
C.~Bohm,
Invent. Math. {\bf 134}, 145 (1998).
\bibitem{ghp2003} 
G.W.~Gibbons, S.A.~Hartnoll, and C.N.~Pope,
Phys. Rev. D {\bf 67}, 084024 (2003).
\bibitem{Lorenz-Petzold1987a} 
D.~Lorenz-Petzold,
Prog. Theor. Phys. {\bf 78}, 969 (1987).
\bibitem{dmpr}
 N. Dadhich, R. Maartens, P. Papadopoulos 
and V. Rezania, Phys. Lett. {\bf B487}, 1 (2000).  
\bibitem{BDW} 
   D. G. Boulware and S. Deser, 
   Phys. Rev. Lett. {\bf 55}, 2656 (1985);
   J. T. Wheeler,
   Nucl. Phys. {\bf B268}, 737 (1986).
\bibitem{bv} 
   M.~Barriola and A.~Vilenkin, 
   Phys. Rev. Lett. {\bf 63}, 341 (1989);
   N.~Dadhich, K.~Narayan, and U.~Yajnik, 
   Pramana {\bf 50}, 307 (1998).
\bibitem{next} 
   H.~Maeda and N.~Dadhich, 
   (to be published).
\bibitem{btz} 
C.~Martinez, C.~Teitelboim, and J.~Zanelli,
Phys. Rev. D {\bf 61}, 104013 (2000).
\bibitem{GBnulldust} 
T.~Kobayashi,
Gen. Rel. Grav. {\bf 37}, 1869 (2005);
H.~Maeda,
Class. Quantum Grav. {\bf 23}, 2155 (2006);
A.E.~Dominguez and E.~Gallo,
Phys. Rev. D {\bf 73}, 064018 (2006).
\bibitem{GBBH} 
   R. -G. Cai, 
   Phys. Rev. D {\bf 65}, 084014 (2002);
   D. L.~Wiltshire,
   Phys. Lett. {\bf B169}, 36 (1986).
\bibitem{dh2006} 
M.H.~Dehghani and S.H.~Hendi,
Phys. Rev. D {\bf 73}, 084021 (2006);
M.H.~Dehghani and R.B.~Mann
Phys. Rev. D {\bf 72}, 124006 (2005).
\bibitem{cai} 
   R.-G.~Cai, 
   private communication.
\bibitem{lz} 
D.~Lynden-Bell and M.~Nouri-Zonoz, 
Rev. Mod. Phys. {\bf 70}, 427 (1998).





\end{references}
\end{document}